\documentclass{iau} 

\newcommand{\bea}{\begin{eqnarray}} 
\newcommand{\eea}{\end{eqnarray}} 

\newcommand{\nn}{\nonumber} 
\newcommand{\vek}[1]{\boldsymbol{#1}}

\renewcommand{\hat}{ { }} 

\usepackage{graphicx}

\title[Gravitational waves from OJ287] %% give here short title %%
{High accuracy measurement of gravitational wave back-reaction in the OJ287 black hole binary}

\author[M.J. Valtonen \& al.]   %% give here short author list %%
{Mauri~J. Valtonen$^{1,2}$, L. Dey$^3$, R. Hudec$^{4,5}$, S. Zola$^{6,7}$, A. Gopakumar$^3$, S. Mikkola$^1$, S. Ciprini$^{8,9}$, K. Matsumoto$^{10}$, K. Sadakane$^{10}$, M. Kidger$^{11}$, K. Gazeas$^{12}$, K. Nilsson$^{2}$, A. Berdyugin$^1$, V. Piirola$^1$, H. Jermak$^{13}$, K.S. Baliyan$^{14}$, D. E. Reichart$^{15}$, S. Haque$^{16}$ \and the OJ287-15/16 Collaboration}

\affiliation{$^1$ Department of Physics and Astronomy, University of Turku, \\ V\"ais\"al\"antie 20,
FIN-21500, Kaarina, Finland \\ email: {\tt mvaltonen2001@yahoo.com} \\[\affilskip]
$^{2}$ Finnish Centre for Astronomy with ESO, University of Turku, \\ V\"ais\"al\"antie 20, FIN-21500 Kaarina, Finland \\[\affilskip]
$^3$ Department of Astronomy and Astrophysics, Tata Institute of Fundamental Research,\\ Mumbai 400005, India\\[\affilskip]
$^4$ Astronomical Institute, The Czech Academy of Sciences, \\25165 Ond{\v r}ejov, Czech Republic\\[\affilskip]
$^5$ Czech Technical University in Prague, Faculty of Electrical Engineering, \\Prague, Czech Republic\\[\affilskip]
$^6$ Astronomical Observatory, Jagiellonian University, \\ul. Orla 171, Cracow PL-30-244, Poland\\[\affilskip]
$^7$ Mt. Suhora Astronomical Observatory, Pedagogical University, \\ul. Podchorazych 2, PL30-084 Cracow, Poland\\[\affilskip]
$^8$ Agenzia Spaziale Italiana (ASI) Science Data Center, \\I-00133 Roma, Italy\\[\affilskip]
$^9$ Istituto Nazionale di Fisica Nucleare, Sezione di Perugia, \\I-06123 Perugia, Italy\\[\affilskip]
$^{10}$ Astronomical Institute, Osaka Kyoiku University, \\4-698 Asahigaoka, Kashiwara, Osaka 582-8582, Japan\\[\affilskip]
$^{11}$ Herschel Science Centre, ESAC, European Space Agency, \\28691 Villanueva de la Ca{\~n}ada, Madrid, Spain\\[\affilskip]
$^{12}$ Department of Astrophysics, Astronomy and Mechanics, National \& Kapodistrian University of Athens, \\Zografos GR-15784, Athens, Greece\\[\affilskip]
$^{13}$ Astrophysics Research Institute, Liverpool John Moores University, \\IC2, Liverpool Science Park, Brownlow Hill, L3 5RF, UK\\[\affilskip]
$^{14}$ Physical Research Laboratory, \\Ahmedabad 380009, India\\[\affilskip]
$^{15}$ University of North Carolina at Chapel Hill, \\Chapel Hill, North Carolina NC 27599, USA\\[\affilskip]
$^{16}$ Department of Physics, University of the West Indies, \\St. Augustine, Trinidad\&Tobago
}

\pubyear{2018}
\volume{338}  %% insert here IAU Symposium No.
\jname{Gravitational Waves Astrophysics: Early Results from Gravitational Wave Searches and Electromagnetic Counterparts}
\editors{Gabriela Gonzalez and Robert Hynes, eds.}
\begin{document}

\maketitle

\begin{abstract}
The binary black hole (BBH) central engine of OJ287 exhibits large thermal flares at least twice every 12 years. The times of these flares have been predicted successfully using the simple rule that they are generated at a constant phase angle of a quasi-Keplerian eccentric orbit. In this model a secondary black hole goes around a primary black hole, impacting the accretion disk of the latter twice per orbital period, creating the thermal flares. New measurements of the historical light curve have been combined with the observations of the 2015/2017 season. The 2015 November/December flare went into the phase of rapid flux rise on the centenary of Einstein's General Relativity, namely on November 25, and peaked on December 5. At that time OJ287 was at its brightest level in over 30 years in optical wavelengths.
 Using the light curve of this flare and subsequent synchrotron flares, and comparing it with the points in the historical light curve, we are able to identify the impact record since year 1886, altogether 25 impacts. Out of these, 14 are timed accurately enough to constrain the orbit of the black hole binary. The set of flare timings determines uniquely the 8 parameters of our BBH central engine model: the two masses, the primary spin, the major axis, eccentricity and the phase of the orbit, plus the two parameters of the standard accretion disk.
 Since the orbit solution is strongly over-determined, its parameters are known very accurately, at better than one percent level for the masses and the primary BH spin. The orbit solution shows that the period of the orbit, now 12.062 year, has decreased at the rate of 36 days per century. This corresponds to an energy loss to gravitational waves that is 6.5 $\pm$ 4 $\%$ less than the rate predicted by the standard quadrupolar GW emission. We show that the difference is due to higher order gravitational radiation reaction contributions to BBH dynamics that includes the dominant order tail contributions. The orbital shrinkage rate agrees within error limits with the rate calculated by Damour, Gopakumar and Iyer (2004).
The full list of participants in the OJ287-15/16 Collaboration is found in ApJL 819, L37, 2016. 

\end{abstract}

\section{Introduction}

OJ287 is a potential nHz gravitational wave source that could be observed by the Pulsar Timing Array method in near future when the detection sensitivity has increased by about a factor of three (Liu et al. 2012, Babak et al. 2016). The nucleus of this galaxy contains a binary black hole system with a 12 yr period and component masses $1.84\times 10^{10}$ $M_{\odot}$ and $1.5\times 10^8$ $M_{\odot}$. Its binary nature was discovered already in 1987 based on huge flares at about 12 yr intervals (Sillanp\"a\"a et al. 1988, see Figure 1). Since then the arrival time of the flares has been predicted successfully in multiple occasions with the rms error of 16 days (See Table 1). In addition, it has been confirmed that the flares belong to the rare category of thermal flares (Valtonen et al. 2012, Valtonen et al. 2016). Therefore it is likely that the flares arise after a collision of the secondary black hole on the accretion disk of the primary (Ivanov et al. 1998, Pihajoki 2016) and it is these collisions that allow us to follow the orbital motion of the secondary around the primary black hole in great detail. The host galaxy of this system has V magnitude about 18 (Takalo et al. 1990, Nilsson 2017) which makes it similar to NGC 4889 in intrinsic brightness; the latter galaxy is one of the two bright central galaxies in the Coma cluster of galaxies, and is believed to harbor a supermassive black hole of mass similar to the primary in OJ287 (Graham and Scott 2013).

\begin{figure}[b]
% \vspace*{-2.0 cm}
\begin{center}
 \includegraphics[width=3.4in,angle=270]{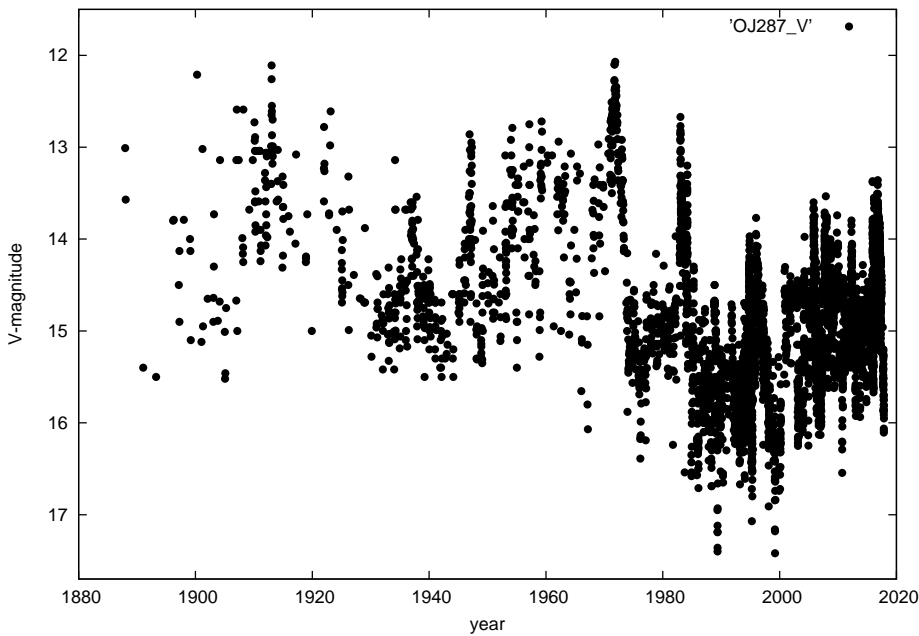} 
% \vspace*{-1.0 cm}
 \caption{Historical light curve of OJ287. It has two dominant periodicities, 12 yr and 56 yr which imply the precession rate of $(12/56)\times180$ degrees per period in the quasi-Keplerian orbit model. The corresponding mass of the primary is $\sim 1.8\times 10^{10}$ $M_{\odot}$. The light variations arise from the combination of varying Doppler boosting due to the wobble of the primary jet (Valtonen and Pihajoki 2013), from varying accretion rate due to tidal forces on the primary accretion disk (Sundelius et al. 1996, Valtonen et al. 2009), and from radiation arising from impacts of the secondary on the accretion disk (Lehto and Valtonen 1996).} 
   \label{fig1}
\end{center}
\end{figure}

\section{Overview}
In recent years a great deal of new data have been added to the optical light curve of OJ287 (Hudec et al. 2013, Hudec 2017). They allow the recognition of 25 major flares since 1886 which may arise from disk impacts. Fourteen of them have good enough light curves to identify the start of the thermal flare (listed in Table 1) while two more are positioned on the time axis by dense observational upper limits (1906 and 1945 flares). Many have occurred during the summer period when OJ287 was not observed (especially in 1920's and 1930's) and cannot be used in this work.
A set of nine flare timings is enough to solve the binary orbit. They determine uniquely the 8 parameters of our BBH central engine model: the two masses, the primary spin, the major axis, eccentricity and the phase of the orbit, plus the two parameters of the standard accretion disk (Valtonen et al. 2010). With the addition of the 2015 flare timing we may explore an additional parameter which in this case is the magnitude of the gravitational wave tail term, not included in previous work. Once a solution for the orbit has been found using ten impact timings, we verify that the solution is consistent with all observed flare times in Table 1, as well as the limits based on the observed upper limits. Thus we have altogether 16 constraints that have to be satisfied by a model with 9 parameters. Obviously, in general such problems do not have a solution, but if a solution is found, it is strongly overdetermined and produces very exact values for the parameters. 
Below we start by describing briefly the gravitational force and gravitational wave model we use. Then we determine the parameters using one hundred orbit solutions and present their mean values and standard deviations around the mean. Finally we discuss the implications of our findings.

\begin{table}
  \begin{center}
  \caption{Overview of quasi-Keplerian binary models of OJ287.}
  \label{tab1}
 \begin{tabular}{|l|c|c|c|c|}\hline 
{\bf flare/} & {\bf 1987-model} & {\bf 1995-model} & {\bf 2006-model} & {\bf 2017-model} \\{\bf parameter}  
   &{\bf [1]} &{\bf [2]} & {\bf [3]} & {\bf [4]} \\ \hline
{\bf1913} & 1913.064 & 1912.124 & 1912.984 & 1912.981 \\ \hline
{\bf1947} & 1948.014 & 1947.304 & 1947.264 & 1947.282 \\ \hline
{\bf1957} & & 1956.004 & 1957.104 & 1957.085 \\ \hline
 {\bf1959}  & 1959.664 & 1959.234 & 1959.184 & 1959.212 \\ \hline
 {\bf1964} & & 1963.794 & 1964.164 & 1964.226 \\ \hline
 {\bf1971} & 1971.314 & 1971.144 & 1971.104 & 1971.127 \\ \hline
{\bf1973} & & 1972.964 & 1972.954 & 1972.927 \\ \hline
{\bf1983} & 1982.964 & 1982.964 & 1982.964 & 1982.964 \\
 &{\bf [5]}&{\bf [5]} &{\bf [5]}&{\bf [5]} \\ \hline
{\bf1984} & & 1984.124 & 1984.124 & 1984.119 \\ \hline
 {\bf1994}  & 1994.614 & 1994.574 & 1994.604 & 1994.596 \\
 &{\bf [6]}& & &{\bf [7]} \\ \hline
 {\bf1995} & & 1995.844 & 1995.824 & 1995.841 \\
&  &{\bf [6]}& &{\bf [8]} \\ \hline
 {\bf2005} & 2006.264 & 2005.704 & 2005.784 & 2005.744 \\ 
& {\bf [6]}& {\bf [6]}& & {\bf [9]} \\ \hline
 {\bf2007} & & 2007.724 & 2007.674 & 2007.691 \\ 
&  &{\bf [6]}& {\bf [6]}&{\bf [8]} \\ \hline
 {\bf2015} & 2017.914 &  & 2015.960 & 2015.875 \\
& {\bf [6]}& &{\bf [6]}&{\bf [10]} \\ \hline
 {\bf primary} & 18.4 & 17.7 & 18.25 & 18.35 \\
$10^9 M_{\odot}$  & & & & \\ \hline
 {\bf secondary} & 0.13 & 0.14 & 0.14 & 0.15 \\ 
 $10^9 M_{\odot}$  & & {\bf [11]} & {\bf [12]} &\\ \hline
{\bf precession} & 38.6 & 33.3 & 39.1 & 38.7 \\ 
 $deg$ & & & &\\ \hline
{\bf spin} & 0 & 0 & 0 & 0.381 \\
 $\chi$ & & & 0.28 {\bf [12]} & \\ \hline
  \end{tabular}
  
 \end{center}
\vspace{1mm}

 {\it Notes:}\\
$^1$ The 1987 tidal model was published in Sillanp\"a\"a et al. (1988). The parameter values for this model given are based on the two dominant frequencies of a quasi-Keplerian binary as determined by Valtonen et al. 2006 and matched with the tidal calculation by Sundelius et al. (1997). $^2$ The 1995 model was solved from timings of the secondary impacting on the accretion disk of the primary, and the time delay between the impact and an observed flare. The disk impact times come from Sundelius et al. (1997) and delay times from Lehto \& Valtonen (1996; LV96), respectively. $^3$ The 2006 model was published in Valtonen (2007). The spin of the primary was assumed zero; a non-zero spin was added in the 2009 model (Valtonen et al. 2010). $^4$ The present 2017 model agrees completely with the observed flare times times. $^5$ All models have been calibrated to the starting time of 1982.964 for the 1983 flare.$^6$ A prediction. The 2015 prediction uses the 2009 model. $^7$ The 1994 flare came within a week from the prediction, considering that the first (thermal) part of the flare was not observable due to closeness of OJ287 to the sun. $^8$ A note in the proofs of LV96 reports the observation of the predicted 1995 flare, exactly as expected. The 2007 flare came within days of the prediction (Valtonen et al. 2008). $^9$ The 2005 flare occurred two weeks ahead of time. $^{10}$ The 2015 flare was known to be spin sensitive. It occurred 4.5 weeks "too early", indicating an increase over the previous spin value. The rms deviation between observations and predictions has been 16 days in five cases. This contrast with the rms deviation for the predictions of the constant period 1987 model, 444 days. Any constant period model would give a similarly poor fit to observations. $^{11}$ Corrected for the Hubble constant H=70 km/s/Mpc. $^{12}$ For the 2009 model.

\end{table}
\section{The Post-Newtonian orbit model}

The relative acceleration between the two black holes
in its center of mass frame, namely $\ddot {\vek x} $, can be divided in several PN contributions (Mora and Will 2004). 
Traditionally, the PN approximation provides the equations of motion
of a binary as corrections to the Newtonian equations of motion
in powers of $(v/c)^2 \sim G M / (c^2 R)$,
where $v$, $M$, $R$ and $c$ are
the characteristic orbital velocity,
the total mass, the typical orbital separation of the binary, and speed of light,
respectively. The relative acceleration and the precessional dynamics of the  spin direction ${\vek s}_1$ are described by
\begin{eqnarray} 
\ddot { {\vek x}} \equiv 
\frac{d^2  {\vek x}} { dt^2} &=& 
\ddot { {\vek x}}_{0} + \ddot { {\vek x}}_{1PN} \nonumber  + \ddot { {\vek x}}_{2PN}
 +  \ddot { {\vek x}}_{2.5PN} \nn\\ 
&& +  \ddot { {\vek x}}_{SS}+ \ddot { {\vek x}}_{3PN}+  \ddot { {\vek x }}_{Q} + \ddot { {\vek x}}_{3.5PN}\\&& + \ddot { {\vek x}}_{4PNtail} + \ddot { {\vek x}}_{SO} +  \ddot { {\vek x}}_{4.5PN} \,,\nn  \\
\frac{d {\vek s}_1}{dt} &=& \left (  {\vek \Omega}_{SO} + {\vek \Omega}_{SS} + {\vek \Omega}_{Q} \right ) \times {\vek s}_1 \,,  
\end{eqnarray} 
where we let  ${\vek x} = {\vek x}_1 - {\vek x}_2 $ as the 
center-of-mass
relative separation vector between the black holes with masses $M_1$ and $M_2$. $  \ddot { {\vek x}}_{0}  $ represents the Newtonian acceleration given by  $ \ddot { {\vek x}}_{0} = 
-\frac{ G\, m}{ r^3 } \, {\vek x} $ where $m= M_1 + M_2$, $ r = | {\vek x} |$. Also we define $ \hat {\vek n} \equiv {\vek x}/r $, $ \dot { {\vek x}} = 
 {\vek v} $ and $\eta = M_1\, M_2/m^2$. 
 We include 
  contributions due to general relativistic  spin-orbt, spin-spin and 
 classical spin orbit interactions, denoted by 
 ${\vek \Omega}_{SO} $, ${\vek \Omega}_{SS} $ and $ {\vek \Omega}_{Q} $, respectively  in the precessional angular velocity 
 for the primary BH spin.
 %refer to orbital angular speed, the angular speed of the secondary and the angular speed of the primary, respectively.

The PN contributions occurring at the conservative 1PN, 2PN, 3PN and the reactive 2.5PN, 3.5PN and 4.5PN orders, denoted 
by $\ddot { {\vek x}}_{1PN}$, $\ddot { {\vek x}}_{2PN}$, $\ddot { {\vek x}}_{3PN}$, 
 $\ddot { {\vek x}}_{2.5PN}$, $\ddot { {\vek x}}_{3.5PN}$ and $\ddot { {\vek x}}_{4.5PN}$, respectively, are non-spin by nature. The explicit expressions for these 
contributions suitable for describing the 
binary black hole dynamics in the modified harmonic gauge are found in Will and Maitra (2017).

The expression for the radiation reaction terms may be written as

\begin{eqnarray} 
\ddot { {\vek x}}_{2.5PN}  =& \frac{8}{5} \frac{ G^2 m^2 \eta }{ c^5 r^3 } 
\biggl \{{A_{2.5}} \dot{r} {\vek n}&   \nn 
- B_{2.5} { \vek v} 
\biggr \} 
, 
\end{eqnarray} 
\begin{eqnarray} 
\ddot { {\vek x}}_{3.5PN}  =& -\frac{8}{5} \frac{ G^2 m^2 \eta }{ c^7 r^3 } 
\biggl \{ A_{3.5} \dot{r} {\vek n}&   \nn 
- B_{3.5} { \vek v} 
\biggr \} 
, 
\end{eqnarray} 
\begin{eqnarray} 
\ddot { {\vek x}}_{4.5PN}  =& \frac{8}{5} \frac{ G^2 m^2 \eta }{ c^5 r^3 } 
\biggl \{ A_{4.5} \dot{r} {\vek n}&   \nn 
- B_{4.5} { \vek v} 
\biggr \} 
, 
\end{eqnarray} 
where, for example, 
%\begin{eqnarray} 
$A_{2.5} = 3 v^2 +  \frac{17}{3} \frac{G\,m}{r} $ and $B_{2.5} =  v^2 + 3 \frac{G\,m}{r}$
%\eea 
%\begin{eqnarray} 
%B_{2.5} =&   v^2 + 3 \frac{G\,m}{r} \nn
%\eea 
while the corresponding expressions for the 3.5 and 4.5 PN orders are more complicated (see Will and Maitra 2017). 
We find that when integrated over a quasi-Keplerian orbit the contribution from the A-coefficients are nearly symmetric but opposite in sign 
with respect to the pericenter. Therefore, they do not contribute much to the orbital averaged quantities. On the other hand, 
the B-coefficients, even though also symmetric with respect to the pericenter, make contributions of the same sign, and are thus important to us. The dominant order hereditary tail contributions to the reactive orbital dynamics are introduced by 
 assuming that they are directly proportional to the ``Newtonian'' radiation term:
\begin{eqnarray} 
\ddot { {\vek x}}_{4PNtails}  =& radfac& \ddot { {\vek x}}_{2.5PN} \nn 
%\,.& 
\end{eqnarray} 
This is mainly due to the absence for closed form expressions for the tail contributions to 
energy and angular momentum fluxes of an eccentric binary (Gopakumar et al. 1997).
The value of $radfac$ is unknown and it is determined as one of the parameters of the orbit solution
and there are on-going investigations to constrain it from theoretical considerations for our BH binary.

The leading order spin-orbit contributions to $ \ddot {{\vek x} }$
appear at the 1.5PN order (Barker \& O'Connell 1979) while the next higher order is at the 2.5 PN level (Will \& Maitra 2017). These terms are included in $  \ddot { {\vek x}}_{SO} $, and they are linear in the Kerr parameter $\chi$. The Kerr parameter and the unit vector 
${\vek s}_1$ define the spin of the primary black hole by the relation
${\vek S}_1 = G\, m_1^2 \, \chi\, {\vek s}_1/c$.
The Kerr parameter $\chi$ is allowed to take values between $0$ and $1$ in GR. 
The terms $ \ddot {{\vek x}_{SS} }$  and $ \ddot {{\vek x}_{Q} }$  include the leading order general relativistic spin-spin interactions
and certain classical spin-orbit interactions that involve the quadrupole moment of the primary BH and they appear 
at the 3PN order for slowly rotating BHs (Will\& Maitra 2017).
%The term $ \ddot {{\vek x}_{Q} }$ includes the leading order contribution of the primary black hole's quadrupole moment and a combination of $(spin)^{2}$ terms and of cross-terms between the quadrupole and monopole terms. These enter at the 3PN level (Will\& Maitra 2017). 
The equations of motion includes certain reactive  4PN spin-orbit contribution, while the $ \ddot {{\vek x}}_{4PNtail}$ term models
the above mentioned leading order gravitational wave tail contributions .
% There is no standard expression for this term, and thus we will include it by using an ambiguity parameter in our orbit solution.

There are 12 degrees of freedom in the $ \ddot {{\vek x}}_{4.5PN}$ term (Gopakumar et al. 1997) . However, these arbitrary parameters completely drop out of the orbit averaged equations for the orbital elements (Will \& Maitra 2017). This allows us to make a free choice of these parameters without significantly affecting the derived orbit, a fact that we have verified by experimenting with different combinations of those parameters. The spin of the primary black hole precesses due to 
the leading order general relativistic spin-orbit coupling 
as described by equation (3.2). The precessional equation for the unit spin vector ${\vek s}_1$ enters the
binary dynamics essentially at the 2PN order.

\section{Solving the orbit}

 The orbit search algorithm starts from a trial orbit, and if the adopted outburst times do not fit, it automatically adjusts itself until a good model orbit is found. There are no solutions in general, but if one is found, it is uniquely defined by the nine parameters of the model. The details of the algorithm are given in Valtonen (2007). 

We find that a solution exits if only the first ``Newtonian'' radiation reaction term (PN2.5) is used. Adding the next level correction (at 3.5PN) leads to a loss of solution, and it is not helped by adding the 4.5PN level. Therefore it is necessary to include the 4PN tail. Since there is no general way to do this for an eccentric orbit, we use an ambiguity parameter $radfac$ as a coefficient of the ``Newtonian'' term and use it as a free parameter in the orbit solution. The adding of an extra parameter is possible since we now have the new tenth fixed point from the timing of the 2015 November flare. 
\begin{table}[h]
 \begin{center}
\caption{Parameters of the orbit solution.
\label{comparisonstars}}
%%\begin{tabular}{lccl}
 \begin{tabular}{|l|c|c|c|}\hline 
Parameter & &unit   & error    \\ \hline 
$M_1$ & 18348 & $10^6 M_{\odot}$&$\pm$101\\ \hline
$M_2$ & 150.13 & $10^6 M_{\odot}$&$\pm$0.43\\\hline 
$\chi_1$ & 0.381 & &$\pm$0.0025\\\hline 
$h$ & 0.845 & &$\pm$0.002\\\hline 
$t_d$ & 0.7734 & &$\pm$0.0015\\\hline
$\Delta\Phi$ & 38.726 & $deg$&$\pm$0.012 \\\hline 
$\Theta_0$ & 55.57  &$deg$ &$\pm$0.21  \\\hline 
$e_0$ & 0.657 & &$\pm$0.001\\\hline 
$P_{2017}$ & 12.062 & yr&$\pm$0.001\\\hline 
$\Delta P$ & 36.2 & d/100 yr&$\pm$0.25\\\hline 
$radfac$ & 1.304 & &$\pm$0.01\\\hline 
$\Delta 2.5PN$ & -0.065 & &$\pm$0.04\\\hline 
$\alpha_g$ & 0.1 & &$\pm$0.05\\\hline 

\end{tabular}
 \end{center}
\end{table}
The fifth column in Table 1 describes the best model with the current data, and the last lines of the same table summarize some of the required parameter values. Most important of the nine independent parameters of the model are the two masses $M_1$ and $M_2$, the spin of the primary $\chi$, and the precession rate of the major axis per period $\Delta\Phi$. Also we determine the eccentricity of the orbit $e_0$, and its orientation at a given epoch $\Theta_0$. The two parameters of the standard accretion disk in the calculation are the disk thickness $h$ and the delay time between an impact and a flare $t_d$, both given in units of Lehto and Valtonen (1996). They translate to more familiar disk parameters $\alpha_g$ (viscosity) and $\dot m $ (mass accretion rate). In addition, we determine the ambiguity parameter related to the 4.5PN term $radfac$. The other properties of interest are the present (redshifted) period of the orbit $P_{2017}$ and the rate of decrease of the period $\Delta P$. From the ambiguity parameter $radfac$ we also derive the correction to the ``Newtonian'' gravitational wave radiation reaction arising from higher order radiation terms $\Delta 2.5PN$. The mean values of these parameters and their standard deviations for 100 hunred solutions are listed in Table 2.    
\begin{figure}[ht] 
\begin{center}
%\epsscale{1.0} 
\includegraphics[angle=270, width=3in] {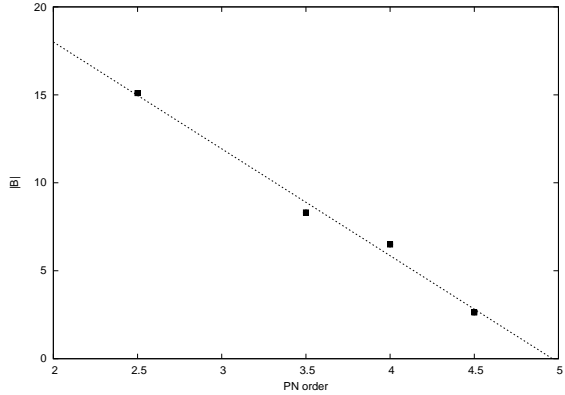} 
%\plotone{Fig59.ps} 
\caption{The absolute value of the B component of gravitational radiation term at the pericenter of the binary orbit, and at different Post Newtonian orders. The sign of the 3.5PN term is opposite of the other terms. Therefore the terms above 2.5PN nearly cancel each other at the pericenter.\label{fig2}} 
\end{center}
\end{figure} 

In Figure 2 the absolute values of the radiation reaction terms and the tail term are compared with each other at the pericenter of the binary orbit. We note that there is a monotonic decrease of the terms with increasing PN order. A linear regression suggests that the 5PN is negligible in the OJ287 problem, and thus we have the required level of accuracy in use. 

\section{Implications}
Since the orbit solution is strongly over-determined, its parameters are known very accurately, at better than one percent level for the masses and the primary BH spin. The orbit solution shows that the period of the orbit, now 12.062 year, has decreased at the rate of 36 days per century. This corresponds to an energy loss to gravitational waves that is 6.5 $\pm$ 4 $\%$ less than the rate predicted by the standard quadrupolar GW emission. We have shown that the difference is due to higher order gravitational radiation reaction contributions to BBH dynamics that includes the dominant order tail contributions. The orbital shrinkage rate agrees within error limits with the rate calculated by Damour, Gopakumar and Iyer (2004).
At present, we are using the system to test General Relativity in the strong field regime unexplored so far. This involves constraining a hypothetical fifth force that arises as a modification of General Relativity in some models of dark matter, dark energy and unification theory. Additionally, we confirm that monitoring of the next flare, predicted to peak on July 31, 2019, at noon GMT, should allow us to test, for the first time, the celebrated black hole no-hair theorem for a massive black hole at the 10$\%$ level (Valtonen et al. 2011).

\end{document}